\PassOptionsToPackage{table}{xcolor}
\PassOptionsToPackage{font={small}}{caption}

\documentclass[sigplan,screen]{acmart}

\copyrightyear{2024}
\acmYear{2024}
\setcopyright{acmlicensed}\acmConference[ASPLOS '24]{29th ACM International Conference on Architectural Support for Programming Languages and Operating Systems, Volume 3}{April 27-May 1, 2024}{La Jolla, CA, USA}
\acmBooktitle{29th ACM International Conference on Architectural Support for Programming Languages and Operating Systems, Volume 3 (ASPLOS '24), April 27-May 1, 2024, La Jolla, CA, USA}
\acmDOI{10.1145/3620666.3651322}
\acmISBN{979-8-4007-0386-7/24/04}

\usepackage[htt]{hyphenat}

\usepackage{tikz}
\usepackage{amsmath}
\usepackage[normalem]{ulem}
\usepackage[loadonly]{enumitem}
\usepackage{booktabs}
\usepackage{multirow}
\usepackage{soul}
\usepackage{listings}
\usepackage{bbding}
\usepackage{comment}
\usepackage[capitalize]{cleveref}
\usepackage[linesnumbered,ruled,lined,noend]{algorithm2e}

\SetCommentSty{mycommfont}
\SetInd{0.5em}{0.5em}
\definecolor{darkgreen}{rgb}{0.0, 0.2, 0.13}
\definecolor{celadon}{rgb}{0.67, 0.88, 0.69}
\definecolor{forestgreen(traditional)}{rgb}{0.0, 0.27, 0.13}
\definecolor{forestgreen(web)}{rgb}{0.13, 0.55, 0.13}
\definecolor{mylightgray}{gray}{0.3}
\SetArgSty{textnormal}

\usepackage{xargs} 

\usepackage{diagbox}

\usepackage{subcaption}

\usepackage{paralist}

\newif\ifsubmit
\submittrue

\ifsubmit
\else
\usepackage[textsize=scriptsize, colorinlistoftodos,prependcaption]{todonotes}
\marginparwidth=\dimexpr \marginparwidth + 1.25cm\relax

\newcommandx{\unsure}[2][1=]{\todo[linecolor=red,backgroundcolor=red!25,bordercolor=red,#1]{#2}}
\newcommandx{\xiangsx}[2][1=]{\todo[linecolor=blue,backgroundcolor=blue!25,bordercolor=blue,#1]{#2}}
\newcommandx{\kun}[2][1==]{\todo[linecolor=orange,backgroundcolor=orange!25,bordercolor=orange,#1]{#2}}

\fi

\newif\ifhidekun
\hidekuntrue

\ifhidekun
\newcommand{\kunhide}[1]{{  }}
\else
\newcommandx{\kunhide}[2][1==]{\todo[linecolor=orange,backgroundcolor=orange!25,bordercolor=orange,#1]{#2}}
\fi

\newcommand{\oursystem}[0]{{Hector}}
\newcommand{\oursystemsmallletter}[0]{{hector}}
\newcommand{\ourtitle}[0]{Hector: An Efficient Programming and Compilation Framework for Implementing Relational Graph Neural Networks in GPU Architectures}

\begin{document}

\title{\ourtitle{}}

\author{Kun Wu}
\email{kunwu2@illinois.edu}
\authornote{Significant portion of the work is done during internship at AWS.}
\orcid{0000-0002-0149-1409}
\affiliation{%
  \institution{University of Illinois at Urbana-Champaign}
  \country{USA}
}

\author{Mert Hidayeto\u{g}lu}
\email{merth@stanford.edu}
\orcid{0000-0001-9276-5075}
\affiliation{%
  \institution{Stanford University}
  \country{USA}
}

\author{Xiang Song}
\email{xiangsx@amazon.com}
\orcid{0000-0001-5030-5054}
\affiliation{%
\institution{AWS AI}
  \country{USA}
}

\author{Sitao Huang}
\email{sitaoh@uci.edu}
\orcid{0000-0001-7669-1467}
\affiliation{%
  \institution{University of California, Irvine}
  \country{USA}
}

\author{Da Zheng}
\email{dazheng2@amazon.com}
\orcid{0000-0001-8115-5415}
\affiliation{%
\institution{AWS AI}
  \country{USA}
}

\author{Israt Nisa}
\email{nisaisrat@amazon.com}
\orcid{0000-0001-5022-5716}
\affiliation{%
\institution{AWS AI}
  \country{USA}
}

\author{Wen-mei Hwu}
\email{whwu@nvidia.com}
\orcid{0000-0003-2532-5349}
\affiliation{%
\institution{Nvidia}
  \institution{University of Illinois at Urbana-Champaign}
  \country{USA}}

\renewcommand{\shorttitle}{An Efficient Programming and Compilation Framework for Implementing RGNNs in GPUs}
\begin{abstract}
Relational graph neural networks~(RGNNs) are graph neural networks with dedicated structures for modeling the different types of nodes and edges in heterogeneous graphs. While RGNNs have been increasingly adopted in many real-world applications due to their versatility and accuracy, they pose performance and system design challenges: inherent memory-intensive computation patterns, the gap between the programming interface and kernel APIs, and heavy programming effort required to optimize kernels caused by their coupling with data layout and heterogeneity. To systematically address these challenges, we propose \oursystem{}, a novel two-level intermediate representation and its code generator framework that 
(a)~\textit{captures} the key properties of RGNN models, and opportunities to reduce memory accesses in inter-operator scheduling and materialization,
(b)~\textit{generates} code with flexible data access schemes to eliminate redundant data copies, and
(c)~\textit{decouples} model semantics, data layout, and operators-specific optimizations from each other to reduce programming effort. 
By building on one general matrix multiply (GEMM) template and a node/edge traversal template, \oursystem{} achieves up to 9.9$\times$ speed-up in inference and 43.7$\times$ speed-up in training compared with the state-of-the-art public systems on select models, RGCN, RGAT and HGT, when running heterogeneous graphs provided by Deep Graph Library~(DGL) and Open Graph Benchmark~(OGB). 
In addition, \oursystem{} does not trigger any out-of-memory (OOM) exception in these tests. 
We also propose linear operator reordering and compact materialization to further accelerate the system by up to 3.8$\times$. 
As an indicator of the reduction of programming effort, \oursystem{} takes in 51 lines of code expressing the three models and generates a total of 8K lines of CUDA and C++ code.
Through profiling, we found that higher memory efficiency allows \oursystem{} to accommodate larger input and therefore attain higher throughput in forward propagation, while backward propagation is bound by latency introduced by atomic updates and outer products.

\end{abstract}



\maketitle

\section{Introduction}

Graph neural networks~(GNNs) have been 
increasingly deployed in various applications, including fraud detection, recommendation, etc.
In response to this growing demand,
the open-source community has made much effort to provide GNN-specific machine learning frameworks, e.g., Deep Graph Library (DGL)~\cite{wang2019deep} and PyTorch Geometric~(PyG)~\cite{fey2019fast}.
These frameworks implement several highly-optimized operations, e.g., sparse-dense matrix multiply~(SpMM) and sampled dense-dense matrix multiply~(SDDMM), to speed up the execution~\cite{huFeatGraphFlexibleEfficient2020a}.
Most of these operators and optimizations are for homogeneous graphs~\cite{huangEfficientSparseMatrix2021,yeSparseTIRComposableAbstractions2022,huFeatGraphFlexibleEfficient2020a}.
However, real-world graphs are typically heterogeneous by nature and contain multiple types of nodes and edges.
For example, a citation graph may represent entities involving authors, articles, etc., as nodes of different types;
the edges may model various types of relations, e.g., an article citing the others.
Recently, to incorporate the information provided by such heterogeneity,
relational GNNs~(RGNNs)~\cite{rgcn,hgt} are proposed to define dedicated parameters and data paths for each type.

RGNN poses {three major} challenges to the existing GPU computation stack due to its inherent computation patterns, the gap between the programming interface and the kernel APIs, and the high cost of kernel code optimizations due to its coupling with data layout and heterogeneity.

The first challenge with GNN implementations on GPUs stems from their need to traverse graphs and {scatter/gather tensor data} in order to use high-performance general matrix multiply~(GEMM) kernels to implement message passing.
In RGNN, message passing is the procedure in each layer where an edgewise operation is followed by a nodewise aggregation operation. In other words, messages are passed through edges to the destination nodes. We show how message passing works in models in Section~\ref{sec:rgnn_formulation}.
During message passing, the graph structure and data layout significantly impact the memory access patterns and execution throughput~\cite{wangEmpiricalAnalysisPerformance2021, zhengNatureGraphNeural2021}. (Examples and details are in Section~\ref{sec:design}).
Furthermore, as the connectivity of the input graph is involved in the {gather} computation, the computation patterns of GNNs are affected not only by the model definition but also by the graph. Such data-dependent behavior precludes any one-size-fits-all optimization strategy when executing GNNs. Additionally, RGNN introduces new complications into the design space due to {the need for the operations to account for heterogeneity. We detail this in Section~\ref{sec:background}.}

The second challenge in RGNN implementation stems from the lack of an abstraction layer between the programming interface and kernel APIs, resulting in extra data movement.  A typical example is an edgewise typed linear layer.
We detail the context and cause of the extra data movement in the edgewise typed linear layer in Section~\ref{sec:segmentmm}. 
But essentially, an edgewise typed linear layer multiplies one of the vectors on each edge with the layer weight dedicated to the edge type.
To achieve this, many existing Pytorch-based systems materialize a temporary three-dimensional edge-wise weight tensor, where the slice corresponding to each edge is the weight matrix of its edge type.
This temporary weight tensor is huge, causing redundant data access and memory footprint.
\oursystem{} avoids such unnecessary copying activities by having typed linear transformations operate on views of tensors, a feature that PyTorch lacks, and decouples the materialization of its operands 
from the source-level expression (Section \ref{sec:materialization}).

Third, code generation is necessary. 
High-performance neural network implementations have historically been based on pre-built libraries, e.g.,  cuBLAS~\cite{nvidiaCublasgemmBatchedCuBLASDocuemntation}. 
GNNs make this less practical because the number of kernels to optimize is multiplied by the number of adjacency-matrix storage format choices such as Blocked-Ellpack~\cite{PMPP4}.
For instance, cuSPARSE only supports the latter in a few APIs~\cite{AcceleratingMatrixMultiplication}.
The typed edges and nodes of RGNN further exacerbate the problem, which makes the traditional pre-built libraries even less adequate and compels framework developers to either painstakingly develop optimized layers from scratch or settle for slow implementation.
For example, it took more than a month for a full-time engineer to implement and deploy the typed linear layer of RGNN in DGL~\cite{nisaFeatureGatherMm}.
Another consequence is the performance degradation caused by limited kernels due to high implementation costs. For example, the  DGL  \texttt{HeteroConv} operator uses a Python native loop to separately launch kernels for each of the relation types in a heterogeneous graph, leading to serial execution of small GPU kernels that underutilize GPU resources on small graphs.

To systematically address these challenges, we propose \oursystem{}, a two-level intermediate representation~(IR) and an associated code generator framework. 
The higher-level IR, called inter-operator level IR, defines the model semantics as sets of operators and expresses layout choices in a decoupled manner. At the lower level, the intra-operator level IR provides the facility to express template specialization and lower them to CUDA kernels. 
We further propose two optimizations, i.e., compact materialization (Section~\ref{sec:materialization}) and linear operator reordering (Section~\ref{sec:inter_op_opt}).
We show in the corresponding Sections how these two optimizations are conveniently enabled by the two-level IR design.
\cref{sec:inter_op_ir,sec:intra-op-ir,sec:ir_design} further the design and rationale of the two-level IR.

In short, \oursystem{} 1)~represents the key properties of RGNN models to capture opportunities to reduce memory accesses in inter-operator scheduling and materialization, 2)~generates code flexibly with proper data access schemes to eliminate redundant data movement, and 3)~expresses model semantics, data layout, and operator-specific optimization in a decoupled manner to reduce programming effort. To the best of our knowledge, \oursystem{} is the first to propose a multi-level IR to capture RGNN-specific opportunities involving cross-relation inter-operator optimizations and tensor data layout with consideration of the type dimension added by RGNNs.
The contribution of this work is as follows.
\begin{enumerate}[1.]
\item We propose the \oursystem{} two-level IR and code generation framework to systematically optimize and generate GPU kernels for RGNN training and inference. It bridges the gap between the programming interface and the kernel generation process, decouples models, data layout, and operator-specific schedule from each other, and leverages optimization opportunities from the three aspects.
\item \label{contrb:eval} We devised the \oursystem{} code generator based on two generalized CUDA templates, {i.e., a} GEMM template and a node and/or edge traversal template. The generated code {achieves} up to 9.9$\times$ speed-up in inference and up to 43.7$\times$ speed-up in training compared to the {best among the} state-of-the-art systems~\cite{wuSeastarVertexcentricProgramming2021, xieGraphilerCompilerGraph, guiHGLAcceleratingHeterogeneous} when running RGCN, RGAT, and HGT~\cite{rgcn, busbridge2019relational, hgt} on heterogeneous datasets provided by DGL and Open Graph Benchmark (OGB) packages~\cite{huOpenGraphBenchmark2021, aifb, mutag, bgs, am, toutanovaObservedLatentFeatures2015}. \oursystem{} also encountered fewer out-of-memory~(OOM) errors, which is significantly alleviated by the optimization mentioned in Section~\ref{contrib:dse}. 
{In fact, with compaction enabled, \oursystem{} incurs no OOM error for all the datasets tested in this paper.}
\item \label{contrib:dse} We devised two optimizations: compact tensor materialization and linear operator reordering.
The best combination of options varies across models and datasets and further obtains up to 3.8$\times$ speed-up in inference and 2.7$\times$ speed-up in training compared to our basic generated code mentioned in Contribution~\ref{contrb:eval}. 
Through profiling, we found that the improved memory efficiency allows \oursystem{} to accommodate larger computation and improve GPU hardware utilization for forward propagation. In contrast, backward propagation does not benefit from larger input, due to its latency-bound nature caused by atomic updates and outer products. 
\end{enumerate}

Artifacts are available at \url{https://github.com/K-Wu/HET}.

\section{Background and Motivation}\label{sec:background}

\subsection{RGNN Formulation and Operators}
\label{sec:rgnn_formulation}
GNNs are feed-forward neural networks that propagate and transform features using the connectivity of a graph. 
A widely used model is graph convolutional network~(GCN)~\cite{kipf2016semi}.
Formally, a GCN layer is defined as 
$    \overrightarrow{{h}^{(l+1)}} = \sigma\left(A^{*}\overrightarrow{{h}^{(l)}}W^{(l)}\right)$
, where $W^{(l)}$ denotes a trainable weight matrix of the $l$-th layer, $\sigma$ is an activation function and $\overrightarrow{{h}^{(l)}}$ is the $l$-th layer node representation. $A^{*}$ is the adjacency matrix normalized by node degrees. $\overrightarrow{h^{(0)}}$ denotes the node input features.

\begin{figure}[!b]
\centering{\includegraphics[width=\linewidth]{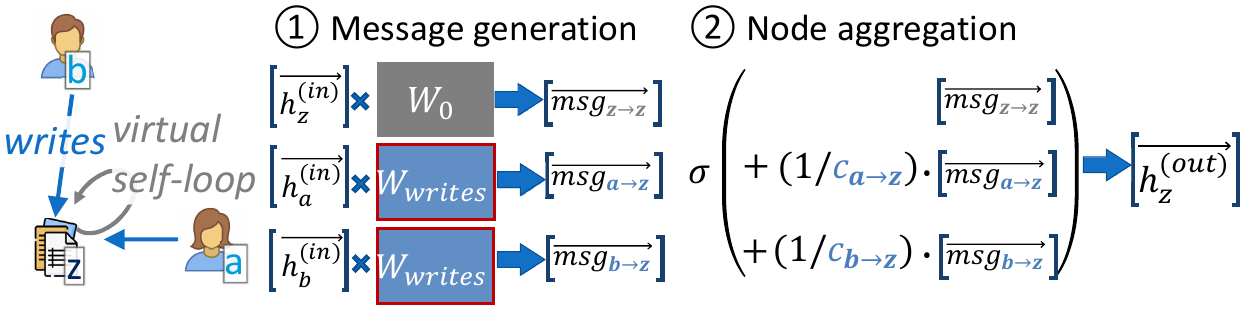}}
\caption{\label{fig:rgnn_layer} The forward propagation of an RGCN layer could be divided into \textcircled{1} message generation on edges and \textcircled{2} node aggregation. We focus on paper node $z$ in a large citation graph as an example. $z$ only has two incoming edges, from $a$ and $b$, respectively. $\overrightarrow{{h}^{(in)}}$ and $\overrightarrow{{h}^{(out)}}$ are node features. $W_{writes}$ is the weight for the type ``writes''. $W_{0}$ is the weight for virtual self-loops. $\sigma$ is the activation function. Notably, some runtime implementations may replicate data, e.g., $W_{writes}$. }
\end{figure}

RGNNs extend GNNs to model different node and edge types for relational graph data.
For example, extended from GCN, a relational graph convolutional network~(RGCN) layer is defined as 
\begin{equation} \label{eq:rgcn}
     \overrightarrow{{h}_v^{(l+1)}} = \sigma\left(\overrightarrow{{h}_v^{(l)}}W_0^{(l)}+\sum_{r \in R}\sum_{u \in \mathcal{N}_v^r}\frac{1}{c_{v,r}}\overrightarrow{{h}_u^{(l)}}W_r^{(l)}\right)
\end{equation}
, where $\mathcal{N}_v^r$ denotes neighbors of node $v$ in relation $r \in R$, 
${h}_n^{(l)}$ is the $l$-th layer node representation of $n$. 
$W_r^{(l)}$ is the weight for relation $r$. $c_{v,r}$ is a problem-specific normalization factor.
Figure~\ref{fig:rgnn_layer} shows an example of how output features is produced in the message passing formulation equivalent to Formula~\ref{eq:rgcn}: 
The forward propagation of an RGNN layer could be divided into \textcircled{1} the edge message generation stage and \textcircled{2} the node aggregation stage.
For simplicity, we focus on the output feature $\overrightarrow{h_z^{(out)}}$ of node $z$: To obtain $\overrightarrow{h_z^{(out)}}$, \textcircled{1} a message $\overrightarrow{msg}$ is generated for each incoming edge, and \textcircled{2} the edge messages go through weighted aggregation and an activation function $\sigma$ to produce $\overrightarrow{h_z^{(out)}}$. 
Notably, to obtain the output feature of node $v$, the input feature of $v$ itself is applied to the $W_0^{(l)}$ and added to the transformed neighbor features. We call this a virtual self-loop because it could be seen as if each node now has a new edge to itself.

\begin{figure}[!bt]
\centering{\includegraphics[width=0.9\linewidth]{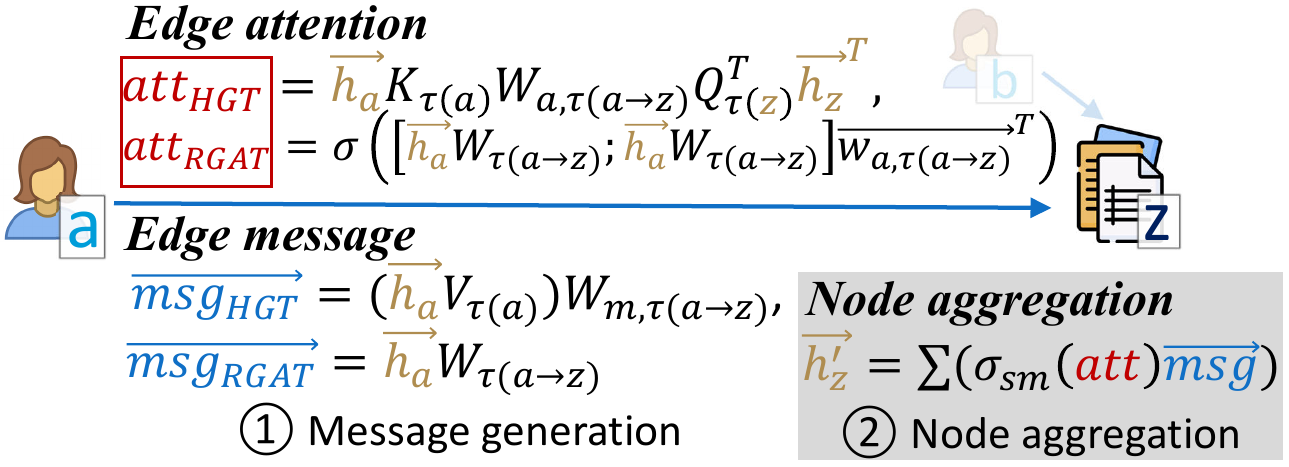}}
\caption{\label{fig:rgat_layer} HGT and RGAT layer. $\overrightarrow{{h}_n}$ and $\overrightarrow{{h}_n^\prime}$ are node $n$'s features. Denote the type of edge from $a$ to $z$ as $\tau(a\rightarrow z)$. Weights $W_{a,\tau(a\rightarrow z)}$ differ by edge type $\tau(a\rightarrow z)$: For example, assuming there are two edge types, ``writes'' and ``cites'', $W_{a,\text{``writes''}}$ is a different weight from $W_{a,\text{``cites''}}$. They are defined and learnt according to the edge type. $W_{m,\tau(a\rightarrow z)}$ and $\overrightarrow{{w}_{a,\tau(a\rightarrow z)}}$ are in similar situations. Weights $W_{\tau(n)}$ differ by the node type $\tau(n)$ of $n$. $\sigma$ is a leaky rectified linear unit (ReLU) in the case of RGAT. $\sigma_{sm}$ stands for edge softmax. $[\vec{s};\vec{t}]$ concatenates $\vec{s}, \vec{t}$.}
\end{figure}

Relational graph attention network~(RGAT)~\cite{busbridge2019relational} and heterogeneous graph transformer~(HGT)~\cite{hgt} are shown in Figure~\ref{fig:rgat_layer}. Attention is introduced in these more complex models:
Attention is produced in the message generation stage together with edge messages.
Similar to the normalization factor, it is a scalar that emphasizes the message associated with the same edge during the subsequent node aggregation stage. However, attention is learned, as is produced by operations among weights and features.

\subsection{RGNN Performance Characteristics}

In {non-graph neural networks,} 
most linear operators, e.g., convolution, can be efficiently implemented with GEMM kernels. 
GEMM takes up most of the execution time due to its cubic complexity.
While some operators can be optimized by transformations, e.g., Winograd for convolution layers~\cite{lavinFastAlgorithmsConvolutional2015}, the operators are still computation-intensive after such computation reduction.
GPUs are excellent at GEMM because the latter's high computation complexity allows leveraging the massive parallel compute units on GPUs, while the input data could be sufficiently reused to allow the memory bandwidth to keep up with the computation throughput.

In contrast, GNNs spend a much larger portion of their execution time on memory-intensive, non-GEMM operations ~\cite{wangEmpiricalAnalysisPerformance2021, zhengNatureGraphNeural2021}. One major source of memory-intensiveness is the sparsity of graphs: to be not bound by the memory {bandwidth}, Nvidia H100 GPU requires the data reuse of single-precision float to be {at least} 16 times. However, the average degree of a graph often falls below this threshold, e.g., the graph datasets in Table~\ref{tab:datasets}. The heterogeneity of RGNNs further exacerbates the issue due to lowered data reuse by the introduction of dedicated weights to different edge types and node types, as shown in Figure~\ref{fig:rgat_layer}.

\begin{figure}[!t]
\centering{\includegraphics[width=\linewidth]{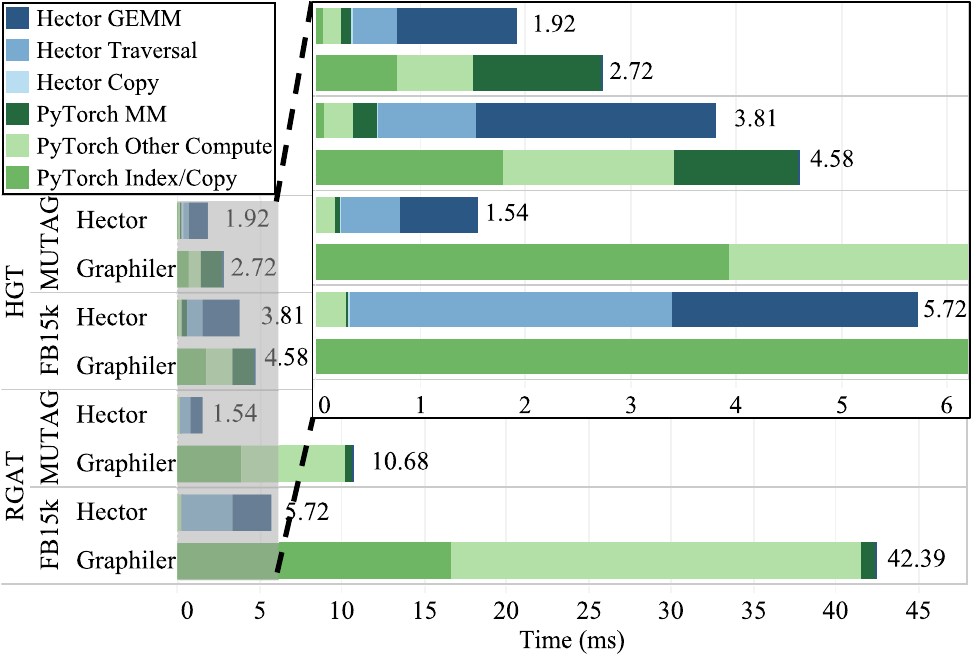}}
\caption{\label{fig:bg_breakdown} Breakdown of inference time by Graphiler and \oursystem{}. Matrix multiply (MM) includes SpMM. We categorize PyTorch time not accounted for by kernels as ``PyTorch Other Compute''.} 
\end{figure}

\begin{figure}[!t]
\centering{\includegraphics[width=\linewidth]{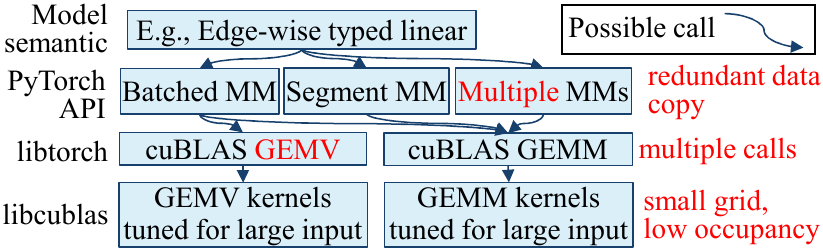}}
\caption{\label{fig:inefficient_stack} Inefficiency (in red) exists in all layers of existing systems.}
\end{figure}

\subsection{Inefficiency in Existing Computation Stack: A Case Study on Edgewise Typed Linear Layers}
\label{sec:segmentmm}

We use an edgewise typed linear layer as an example to walk through the various performance overheads in the existing computation stack, as summarized in Figure~\ref{fig:inefficient_stack}. 
Edgewise typed linear layer applies a typed linear operator on each edge to one of its vectors. The weight of the linear operator used in the computation depends on each edge's type. For example, the edge message in an RGCN layer (Figure~\ref{fig:rgnn_layer}) or an RGAT layer (Figure~\ref{fig:rgat_layer}), is produced by a typed linear layer.

A typed linear layer is typically implemented using batched matrix multiply (BMM) or segment matrix multiply (segment MM)~\cite{isratsegmm}. 
For example, PyG \texttt{FastRGCNConv} implemented typed linear layers using BMM to unleash parallelism. However, a temporary tensor must be created from the weight tensor due to the lack of support for indirect addressing by PyTorch tensor APIs: the typed linear layer could be denoted as  $Y[i,0,j]:=\sum_k(X[i,0,k]\times W[T[i],k,j])$ where $X[i,0,\cdot]$, $Y[i,0,\cdot]$ and $W[T[i],\cdot,\cdot]$ are input feature, output feature of node $i$ and the weight of node $i$'s type. The middle dimension of $X$ and $Y$ are needed to make the operation a matrix multiply. However, there is currently no support for specifying $T[i]$ as one of the arguments to an operator in PyTorch;  
one must create $W^\prime[i,k,j]:=W[T[i],k,j]$ before the typed linear layer.

Segment MM requires presorting features by types. After that,  the node/edge feature tensor is in the form of segments of features of the same type: the segment MM kernel then applies the corresponding weight tensor of the type to each segment. 
If neither BMM nor segment MM can be employed, one may fall back to multiple matrix multiplies, leading to higher device API overhead and GPU under-utilization.

Another type of inefficiency is suboptimal math library calls. PyTorch has routines to handle various scenarios, e.g., a tensor is strided in memory layout or is \texttt{NestedTensor}, a pack of tensors. Consequently, Pytorch sometimes performs BMM by launching multiple general matrix-vector multiplies (GEMVs) kernels, which also leads to API overhead and GPU under-utilization.

Lastly,  CUDA math libraries were initially developed for large inputs and may not be {efficient} for small inputs~\cite{nvidiaCublasgemmBatchedCuBLASDocuemntation}.

To better illustrate the points, Figure~\ref{fig:bg_breakdown} breaks down HGT and RGAT inference time on FB15k and MUTAG.
Section~\ref{sec:eval_methodology} details the system configurations and datasets.
This experiment measured Graphiler~\cite{xieGraphilerCompilerGraph}, which executed compiled TorchScript code and delivered the best inference performance among the existing systems tested in this work.
Figure~\ref{fig:bg_breakdown} shows that indexing and copying take up a significant portion, and the portion of GEMM operations, i.e., MM vs. Other compute,  varied with graphs.
By profiling, we found that the CUDA API overhead is 22\% the time of the critical path, which is the sum of the API overhead and kernel duration. This is partly due to a huge number of kernel launches caused by 1) libraries calling a series of kernels to fulfill an API invocation and 2) some operators calling separate sets of kernels for each types in the graph.

\begin{figure*}[!htbp]
\centering{\includegraphics[width=0.7\linewidth]{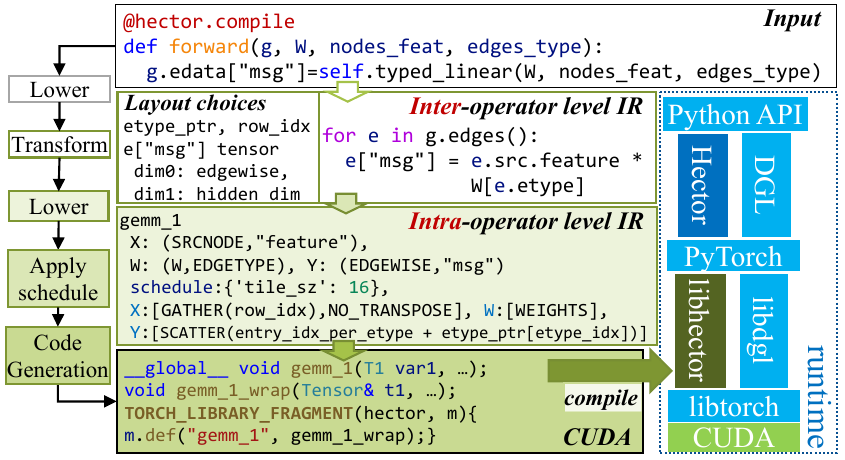}}
\caption{\label{fig:runtime_arch} \oursystem{} workflow and software architecture. }
\end{figure*}

In contrast, \oursystem{} 1) \textbf{lowers more of the logic to GEMM}, 
and 2) assembles kernels with flexible access scheme to \textbf{gather and scatter data on the fly} to eliminate redundant data movement. Consequently, \oursystem{} does not replicate weights during computation. As shown, \textbf{this strategy achieves better performance than using hand-optimized kernels with dedicated functions to data movement, e.g., in Graphiler}.

To address the performance challenges in RGNN systems due to both RGNN's inherent computation pattern and the system design, we propose the \oursystem{} IR and code generation framework. By the IR design that \textit{decouples} and \textit{expresses} the model semantic, data layout, and operator-specific schedules, \oursystem{} opens up these opportunities and the integration of all three aspects into the design space.
Table~\ref{tab:salesman_matrix} shows the feature comparison of \oursystem{} with existing systems. 

\begin{table}[!b]
\caption{Features of \oursystem{} and prior~\cite{xieGraphilerCompilerGraph, wuSeastarVertexcentricProgramming2021, guiHGLAcceleratingHeterogeneous} GNN compilers.\label{tab:salesman_matrix}
}
\centering
{\footnotesize
\begin{tabular}{@{}llllll@{}}
\toprule
&                                & \rotatebox{75}{Graphiler} & \rotatebox{75}{Seastar}   & \rotatebox{75}{HGL}       & \rotatebox{75}{\textbf{\oursystem{}}}   \\ \midrule
\multirow{2}{*}{\textbf{Target}}                                                 & \textbf{Inference}             & \Checkmark & \Checkmark &           & \Checkmark       \\
& \textbf{Training}              &           & \Checkmark & \Checkmark & \Checkmark       \\\cline{1-2}
\multicolumn{2}{@{}l}{\textbf{Memory efficiency}}                          & \Checkmark &           & \Checkmark & \textbf{better} \\\cline{1-2}
\multirow{3}{*}{\textbf{\begin{tabular}[c]{@{}l@{}}Design\\ space\end{tabular}}} & \textbf{Data layout}           &           &           &           & \Checkmark       \\
& \textbf{Intra-operator schedule}     &           &           &           & \Checkmark       \\
& \textbf{Inter-operator optimization} & \Checkmark & \Checkmark & \Checkmark & \Checkmark       \\ \bottomrule
\end{tabular}
}
\end{table}

\section{Design and Implementation of \oursystem{}}\label{sec:design}

\subsection{Overview of Workflow and System Components}
\oursystem{} consists of a programming interface, a code generator, and Python modules.
The code generator takes in the model definition and generates both CUDA kernels and host functions that configure and invoke the CUDA kernels.

Figure~\ref{fig:runtime_arch} uses an example to illustrate the workflow. The input is an excerpt of DGL code invoking a typed linear layer on the input node features. Applying the \texttt{@\oursystemsmallletter{}.compile} decorator triggers a transpiling pass to lower the code into \oursystem{} inter-operator level IR. {In this example, the typed linear transformation \texttt{typed\_linear}} can be efficiently implemented as GEMM kernels. {To this end,} \oursystem{} lowers the transform to an operator instance derived from the GEMM template at the inter-operator level. {After the analysis and optimizations at the inter-operator level, \oursystem{} further lowers the code to a detailed GEMM specification at the intra-operator level.} The GEMM output $A$ collects edge data generated from the node data. The first input $B$ is the weight matrix $W$, and the second input $C$ is the collection of features of all the source nodes of the edges involved. The intra-operator level IR indicates that the GEMM operation should use the default tile width of 16 and be carried out without scatter, gather, or transpose applied to input or output matrices. Eventually, \oursystem{} generates a segment MM~(Section~\ref{sec:segmentmm}) kernel, \texttt{gemm\_1}. 
The Layout Choices section of Figure~\ref{fig:runtime_arch} shows the default layout choice. \texttt{etype\_ptr} specifies the offsets of each segment of different type. \texttt{row\_idx} is the source node index array in the COO format. The result tensor \texttt{e["msg"]} has the number of edges as the number of rows, and the number of the columns is the input dimension of the hidden layer. We detail in Section~\ref{sec:materialization} an optimization technique, compact materialization, that is opened up by the decoupled layout choices from the inter-operator level IR.

The generated code is compiled into a shared library where host functions are exported through the \texttt{pybind11} utilities.
\oursystem{} falls back to existing routines in PyTorch when certain operators are not yet supported.
During runtime, the precompiled functions are loaded and registered as subclasses of PyTorch \texttt{autograd.Function}.

\begin{figure*}[!htbp]
\centering{\includegraphics[width=0.95\linewidth]{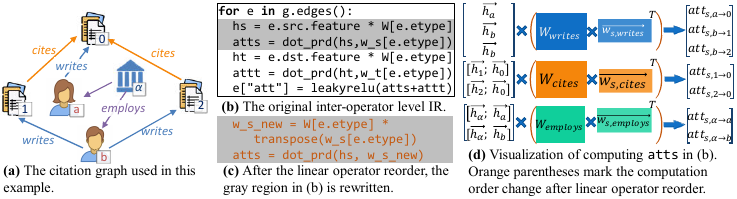}}
\caption{\label{fig:linear_opt} In the example graph in (a), when computing edge attention of RGAT, linear operator reordering could be applied. (b) shows the original inter-operator level IR to compute RGAT edge attention. (d) visualizes the computation of the first term, \texttt{atts}, and uses the orange parentheses to mark how the linear operator reordering changes the order of the computation. (c) The transformation rewrites the code.}
\end{figure*}

\begin{figure*}[!htbp]\captionsetup[subfigure]{font=small}
\centering
\subcaptionbox{GEMM kernel and IRs of RGAT edge message computation with vanilla materialization. The two red squares mark identical terms because \texttt{msg} depends only on source node and edge type. Both schemes in (a) and (b) \textcircled{1} gather the source node’s features into a matrix, \textcircled{2} perform the GEMM computation, and \textcircled{3} scatter the output features to rows in the output tensor. Each dotted square mark a block in \textcircled{2} the GEMM kernel. \texttt{row\_idx} specifies the source node index of each edge, and is used in step \textcircled{1}. \texttt{etype\_ptr} specifies the offsets of edge of each type and is used in step \textcircled{3}.}
[\linewidth]{\includegraphics[scale=1.4]{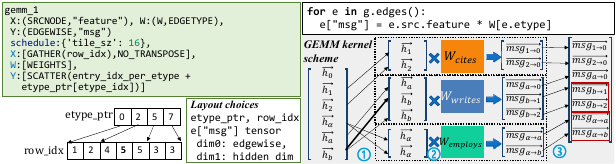}}
\subcaptionbox{GEMM kernel and IRs of RGAT edge message computation with compact materialization. Differences in IRs are marked in orange.}
[\linewidth]{\includegraphics[scale=1.4]{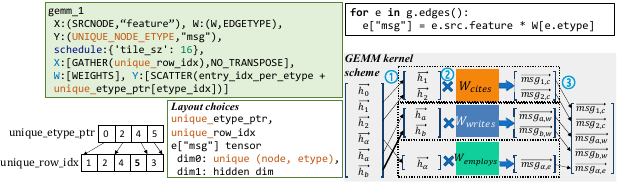}}
\caption{\label{fig:compact_opt_opt} When computing RGAT edge messages, compact materialization could be applied. This figure uses the graph in Figure~\ref{fig:linear_opt}(a). Compared with (a) vanilla materialization, (b) compact materialization saves both the memory footprint and the computation. In (a), the leading dimension of the output message tensor accommodates different edges. In (b), it accommodates unique (source node, edge type) pairs. \texttt{unique\_row\_idx}, and \texttt{unique\_etype\_ptr} describes the mapping from (source node index, edge type index) to the unique index.}
\end{figure*}

\subsection{Inter-Operator Level IR}
\label{sec:inter_op_ir}

The inter-operator level IR follows the Python grammar but involves some new constructs, as listed in Table~\ref{tab:ir_constructs}. Listing~\ref{lst:ir_example} illustrates how the attention calculation in a single-headed RGAT layer could be expressed using the inter-operator level IR.
Lines 10-16 shows a code segment that generates attention values for all edges of graph \texttt{g} and then invoke the \texttt{edge\_softmax(g)} function that spans lines 1 through 9. As shown in Listing~\ref{lst:ir_example}, the message generation and aggregation stages are expressed as for-each edge loops starting from line~2, line~8 and line~10, and for-each node loop starting from line~4. To accumulate data from the incoming edges of each node n, the \texttt{n.incoming\_edges()} iterator is used. Notably, the data layout that specifies how to access the input and output data per edge or node as well as the incoming edges associated with each node, is abstracted away in Listing~\ref{lst:ir_example}.

\subsubsection{Programming Interface}
\oursystem{} provides a decorator, \texttt{@\oursystemsmallletter{}.compile}, to take the existing PyG or DGL forward propagation logic and generate code for it, as exemplified by the input in Figure~\ref{fig:runtime_arch}. The decorator, when applied to a method, invokes a simple transpiling pass that replaces the PyG and DGL method calls, e.g., SpMM/SDDMM,  with an implementation in the inter-operator level IR, and replaces supported constructs from PyG and DGL with expressions in \oursystem{} IR.
Similarly to statically-typed compilers in other Python packages~\cite{pytorchTorchScriptPyTorchDocumentation, lamNumbaLLVMbasedPython2015}, the function to compile can use most of the Python features except dynamic ones, e.g., assigning objects of different types to the same variable. We support a few types as the function arguments for heterogeneous graphs, involving \texttt{Tensor} and \texttt{dict[str, Tensor]} objects, i.e., \texttt{dict} objects where the keys are \texttt{str} objects and the values are \texttt{Tensor} objects.

Besides, one can use the \oursystem{} inter-operator level IR itself to express the model, as exemplified by Listing~\ref{lst:ir_example}. %

\begin{lstlisting}[caption={Expressing the attention calculation in a single-headed RGAT model using \oursystem{} inter-operator level IR.},label={lst:ir_example},language=Python]
def edge_softmax(g):
    for e in g.edges():
        e["att"] = exp(e["att"])
    for n in g.dst_nodes():
        n["att_sum"] = 0.0
        for e in n.incoming_edges():
            n["att_sum"] += e["att"]
    for e in g.edges():
        e["att"] /= e.dst["att_sum"]
for e in g.edges():
    hs = e.src.feature * W[e.etype]
    atts = dot_prd(hs, w_s[e.etype])
    ht = e.dst.feature * W[e.etype]
    attt = dot_prd(ht, w_t[e.etype])  
    e["att"] = leakyrelu(atts + attt)
edge_softmax(g)
\end{lstlisting}

\subsubsection{Compact Tensor Materialization and Data Layout}
\label{sec:materialization}

{The} \oursystem{} inter-operator level IR deliberately abstracts away the data layout from the model semantics. As exemplified by Listing~\ref{lst:ir_example}, the IR only expresses the association of variables with nodes or edges, e.g., \texttt{e["att"]} and \texttt{n["att\_sum"]}, without dictating the mapping of elements in the conceptual variable to the memory space. 

In this work, we devised compact materialization, which is a technique enabled by the decoupling between model semantics and data layout. 
Note that certain edge data are determined by sparse combinations of source node features and edge types, e.g.  $\overrightarrow{{msg}_{HGT}}$ in Figure~\ref{fig:rgat_layer}. Rather than computing and storing such data for each edge, we instead compute and store the data once for each $\left(\text{edge type}, \text{unique node index}\right)$ pair  that actually exists, reducing the resources spent on computing and storing common subexpressions.
As exemplified in Figure~\ref{fig:compact_opt_opt}, the materialized tensor involves seven rows when each row vector corresponds to a \texttt{msg} of an edge.
Alternatively, the system can materialize the tensor with only five rows, where each row vector corresponds to a \texttt{msg} of an $\left(\text{edge type}, \text{unique node index}\right)$ pair.
We call the former vanilla materialization and the latter compact materialization.
For the vanilla scheme, the row number is the edge index specified by the sparse adjacency. For the compact scheme, it is a unique non-negative integer assigned to each $(\text{source node}, \text{edge type})$. We precompute this mapping and store it in a CSR-like format. \oursystem{} does not create the temporary weight tensor, as explained in Section~\ref{sec:segmentmm}.
In summary, compact materialization is a technique to eliminate repetitive identical computations and results in edgewise operators. It is applicable when an edgewise operator depends only on the source node data and edge type, and its output has the shape of $(\text{number of edges}, \text{hidden dimension size})$. After this optimization, the output shape is reduced to $(\text{number of unique} \allowbreak (source\ \allowbreak node, edge\ type)\text{ pairs}, \text{hidden dimension size})$, and repetitive computation is eliminated.
Section~\ref{sec:dse_eval} provided further analysis of the effects of compact materialization on memory footprint reduction.

\begin{table}[!htbp]
\centering
\caption{\oursystem{} inter-operator level IR constructs. The graph's variable is named as \texttt{g}, node's as \texttt{n}, and edge's as \texttt{e}. }\label{tab:ir_constructs}
{\footnotesize
\begin{tabular}{@{}l@{}l@{}l@{}l@{}}
\multicolumn{4}{l}{\textbf{Methods of graph variables}}\\\hline
\multicolumn{1}{|l}{node iterator}        & \multicolumn{3}{l|}{\texttt{g.dst\_nodes()}, \texttt{g.src\_nodes()}}                                                      \\\hline
\multicolumn{4}{|l|}{
\noindent{
\begin{tabular}[t]{@{}ll|ll@{}}
\hspace{-2.25pt}edge iterator        & \texttt{g.edges()}  & weight slicing, e.g.,    &\texttt{W[e.etype]} \\
\end{tabular}}}     \\\hline
\multicolumn{1}{|l}{neighbor iterator}    & \multicolumn{3}{l|}{\texttt{n.incoming\_edges()}, \texttt{n.outgoing\_edges()}}\\\hline 
\multicolumn{4}{l}{\rule{0pt}{10pt}\textbf{Attributes}}\\\hline
\multicolumn{4}{|l|}{
\noindent{
\begin{tabular}[t]{@{}ll|ll@{}}
\hspace{-2.25pt}nodes                & \texttt{e.src}, \texttt{e.dst}                       & types                       & \texttt{e.etype}, \texttt{n.ntype}  
\end{tabular}}}     \\\hline
\multicolumn{4}{|l|}{
\noindent{
\begin{tabular}[t]{@{}ll|ll@{}}
\hspace{-2.25pt}input data, e.g.,    & \texttt{n.feature}      &produced data, e.g.,& \texttt{e["att"]}
\end{tabular}}}     \\\hline
\multicolumn{4}{l}{\rule{0pt}{10pt}\textbf{Operators}}\\\hline
\multicolumn{2}{|l}{GEMM-eligible computation, e.g.,}        & \multicolumn{2}{l|}{\texttt{linear()}, \texttt{outer\_prod()}}                                                      \\\hline
\multicolumn{2}{|l}{GEMM-ineligible computation, e.g.,}    & \multicolumn{2}{l|}{\texttt{dot\_prod()}} \\\hline
\multicolumn{2}{|l}{manipulation, e.g.,}    & \multicolumn{2}{l|}{\texttt{reshape()}, \texttt{concat()}} \\\hline
\end{tabular}
}
\end{table}

Besides tensor materialization, the multi-level IR design also allows data layout optimizations involving 1) architecture-specific optimizations, e.g., padding, and 2) various sparse adjacency encoding.
At the inter-operator level, data layout specifications are decoupled from the model semantics and do not influence the transform passes at this level. 
However, they determine the data access scheme and make a difference when generating CUDA code at the intra-operator level.
\oursystem{} inter-operator level IR bookkeeps the specifications, which are passed to the intra-operator level during lowering. 
The intra-operator level operator instances choose the data access scheme corresponding to the data layout specifications while assembling the kernel code.
We leave the exploration of data layout optimizations to future work and detail our plan in Section~\ref{sec:future_work}.

\subsubsection{Linear Operator Reordering}
\label{sec:inter_op_opt}
Linear operator reordering is an inter-operator level optimization. When a linear operator, e.g., linear layer and dot product, is followed by another linear operator, their order may be switched. 
For example, for \texttt{atts} as shown in Figure~\ref{fig:linear_opt}(d), we may calculate $W_r\vec{w}_{r}^T$ first instead. Its profitability can be determined by counting the number of multiplication and addition involved in the two GEMMs before and after the order is changed. For now, we implement the pass to switch the orders of two linear operators whenever this produces an operator between weights, because it reduces the complexity by reducing one of its factors, the number of nodes/edges, to the size of hidden dimension. For simplicity, rewritten operator instances use PyTorch BMM to compute the product of weights and apply PyTorch slicing when necessary.

\subsubsection{Graph-Semantic-Aware Loop Transformation}\label{sec:graph_aware_loop}
Loop transformation at this level is augmented with the graph-semantic-specific equivalence rule: a for-each loop over the edges is equivalent to a for-each loop nest iterating over all the incoming/outgoing edges of all destination or source node{\color{red}}. Loop transformation is applied during the lowering pass to canonicalize and fuse loops in order to more thoroughly identify kernel fusion opportunities.

\subsubsection{Lowering Inter-Operator Level IR}

To lower the IR to the intra-operator level, \oursystem{} greedily lowers every eligible operator to instances derived from GEMM templates (Section~\ref{sec:two_templates}). Then, it fuses each remaining region and lower them to as few traversal instances (Section~\ref{sec:two_templates}) as possible.
To achieve this, \oursystem{} scans the code three times. Each time, it attempts to lower operators to instances of a specific preference level. During the first pass, it attempts to lower operators to GEMM-template-derived instances. In the next pass, it attempts the traversal-template-derived instances. The third pass will lower all the remaining operators to PyTorch function calls.
During each pass, whenever an operator can be lowered, \oursystem{} marks the operator itself, together with all subsequent operators that can be fused into it, with the lowering decision. 
After all the operators have been examined in a pass, the marked operators are lowered and fused. Before the second pass, it canonicalizes the for loops and fuses loop nests whenever possible to discover kernel fusion opportunities.

\subsection{Intra-Operator Level IR}\label{sec:intra-op-ir}

{The} intra-operator level IR serves between the inter-operator level IR and the generated CUDA code. At this level, the IR should encode specifications to emit CUDA code and provide sufficient information specific to each operator invocation to the transform and lowering passes at the inter-operator level. 
The code transformation components at this level provide the methods to generate specialized CUDA code for the operators, to apply operator-specific schedules, and to return necessary information on operator selection and kernel fusion feasibility to the passes at the inter-operator level.

\oursystem{}'s code generator ultimately lowers the IR to two basic constructs, the GEMM template and the traversal template. 
Algorithms~\ref{algo:gemm_template} and~\ref{algo:traversal_template} illustrate the edge traversal template and the GEMM template. 
The node traversal template is similar to Algorithm~\ref{algo:traversal_template}, and we will revisit it in Section~\ref{sec:op_schedule}.
For simplicity, function template specialization refers to routines specialized for the specific instances derived from the two templates and involve 1)~function arguments, e.g., number of rows, etc., 2)~special registers, e.g., \texttt{threadIdx}, and 3) loop variables.

\subsubsection{The GEMM Template and the Traversal Template}\label{sec:two_templates}
We base the code generation on GEMM and traversal templates because RGNNs involve not only sparse operations but also multiple dense operations to project vectors across different semantic spaces.
The GEMM template serves edgewise and nodewise linear transformations, as exemplified by the computation of RGAT edge messages in Figure~\ref{fig:compact_opt_opt}. The GEMM template is defined as a matrix multiply augmented with custom gather and scatter schemes. It is formulated as $Y[S] = X[G] \times W[T]$ where $Y$, $X$, $W$ are output, input, and weights, respectively; $S$, $G$ and $T$ are scatter list, gather list, and the type of the nodes or edges, respectively. 
The traversal template performs generic nodewise or edgewise operations. It serves operators that cannot be lowered to GEMM templates, e.g., edgewise dot products.

As shown in Algorithm~\ref{algo:gemm_template}, the GEMM template is based on tiled matrix multiplication. The GEMM template starts with %
the work assignment per block during the \texttt{GetRange<kid>} subroutine (line 1). 
The \texttt{idxTileRow} and \texttt{idxTileCol} whose range is determined by \texttt{GetRange<kid>} is used to position the workload.
Typically, it is the coordinate of the tile of the output matrix.
Factors that affect $X$'s loading scheme, \texttt{LoadXToShmemIfInRange<kid>}, and $W$'s, \texttt{LoadWToShmemOrRegistersIfInRange<kid>}, involve whether gather lists or transpose needs to be applied on the fly (lines 4-5).
Gather list $G$ in the Input section is sometimes needed to locate the rows in the source matrix $X$: For example, in Figure~\ref{fig:compact_opt_opt} (a), \texttt{row\_idx} is needed in step \textcircled{1}.
The required information will be passed during the lowering.
The operator instance then accordingly chooses the data access scheme code piece for kernel code generation.
The storing scheme \texttt{StoreCIfInRange<kid>} depends similarly on whether a scatter list will be applied. 
Atomic intrinsics are used in the case of multiple simultaneous updaters.

In the traversal template, as shown in
Algorithm~\ref{algo:traversal_template}, the edge type, node indices retrieval scheme in lines~5-7 depend on the sparse adjacency encoding.
Similarly to the GEMM template, when a row vector needs to be loaded or stored, the tensor materialization scheme determines how the row is located in the materialized tensor.
All statements are initially inserted into the innermost loop. 
After \oursystem{} finishes the loop transformations, it then defines work assignment on line~1 in Algorithm~\ref{algo:traversal_template} for the operator instance derived from the traversal template using a simple scheme. For example, if the loop nest is three levels, as exemplified by Algorithm~\ref{algo:traversal_template}, we assign the outermost loop, i.e., \texttt{idxEdge} or \texttt{idxNode} loop, to each thread block and the two inner loops to the multi-dimensional threads in each block.

\begin{algorithm}[!htbp]
{\small
\KwIn{References of Tensor $Y, X, W$, gather list $G$, etc.}
\texttt{tileRowRange}, \texttt{tileColRange} $\gets$ \textbf{\texttt{GetRange<kid>}}()\;
\ForEach{\texttt{idxTileRow} $\in$ \texttt{tileRowRange}}{
\ForEach{\texttt{idxTileCol} $\in$ \texttt{tileColRange}}{
\textbf{\texttt{LoadXToShmemIfInRange<kid>}}()\;
\textbf{\texttt{LoadWToShmemOrRegistersIfInRange<kid>}}()\;
\texttt{\_\_syncthreads()\;}
\texttt{Y\_reg} $\gets$ \texttt{X\_shmem} $\times$ \texttt{W\_shmem\_or\_reg}\;
\texttt{\_\_syncthreads();}
}
\textbf{\texttt{StoreYIfInRange<kid>}}()\;
}
}
\caption{\small\oursystem{}'s GEMM template in pseudo-code. Each instance is assigned a unique identifier \texttt{kid} and gets function template specialization \textbf{\texttt{FuncName<kid>}}.}
\label{algo:gemm_template}
\end{algorithm}

\begin{algorithm}[!htbp]
{\small
\KwIn{References of input and output tensors. Other necessary data, e.g., adjacency.}
\texttt{eRange}, \texttt{hRange}, \texttt{fRange} $\gets$ \textbf{\texttt{GetRange<kid>}}()\;
\ForEach{\texttt{idxEdge} $\in$ \texttt{eRange}} {
    \ForEach{\texttt{idxHead} $\in$ \texttt{hRange}}{
        \ForEach{\texttt{idxFeat} $\in$ \texttt{fRange}}{
        \texttt{eType} $\gets$ \textbf{\texttt{GetEType<kid>}}()\;
        \texttt{srcIdx} $\gets$ \textbf{\texttt{GetSrcId<kid>}}()\;
        \texttt{dstIdx} $\gets$ \textbf{\texttt{GetDstId<kid>}}()\;
        \tcp{initial insertion point}
        }
    }
}
}
\caption{\small\oursystem{}'s edge traversal template in pseudo-code. Similarly to Algorithm~\ref{algo:gemm_template}, each instance gets specialized \textbf{\texttt{FuncName<kid>}}.}
\label{algo:traversal_template}
\end{algorithm}

\subsubsection{Adapting to Different Sparse Adjacency Encoding}
At the intra-operator level, the templates work for any sparse adjacency encoding as long as specific interfaces are implemented. For example, the edge traversal shown in Algorithm~\ref{algo:traversal_template} works as long as the function template specialization \texttt{GetEType<kid>}, \texttt{GetSrcId<kid>} and \texttt{GetDstId<kid>} are implemented: If the sparse adjacency is COO, \texttt{GetSrcId<kid>} is a subscript operator applied to the row indices array. If it is CSR, then \texttt{GetSrcId<kid>} is a binary search in the row pointer array.

\subsection{Rationale of the \oursystem{} Two-Level IR}
\label{sec:ir_design}

Central to the code generator is the two-level IR.
Inter-operator level IR optimizations address the opportunities brought in by heterogeneous relation types. These optimizations manipulate operators and their connections. A high-level IR %
abstracts away the low-level details that can complicate or even hinder the transformations. 
Intra-operator level IR optimizations reduce the data movement by generating access schemes in kernels rather than using specialized kernels and dedicated indexing/copying kernels. These optimizations manipulate low-level data access and schedule details, and thus are better supported by a low-level IR.

The two-level IR enables concerted but decoupled choices of intermediate data layout and compute schedules:
For example, in Figure~\ref{fig:runtime_arch}, the semantics of the model are decoupled from the layout choices.
\oursystem{} implements the model semantics and layout choices in intra-operator level IR with specific access schemes.
The next few paragraphs explain how the two-level IR design facilitates operator-specific optimizations, operator selection, and kernel fusion.

\subsubsection{Operator-Specific Schedule}
\label{sec:op_schedule}
Each instance derived from the GEMM template provides the option to apply a coarsening factor in $\{2,4\}$, to choose the tile size, and to apply \texttt{\_\_launch\_bounds\_\_} that limits the number of registers in exchange for more active warps. 
The coarsening factor is the number of elements each thread deals with in the loading, computing, and storing stages. When applied, each block still works on the same assignment, but its number of threads shrinks by the factor~\cite{PMPP4}.
We also allow a per-row scalar to be applied to the tiles of matrix $A$.
This eliminates the extra memory-intensive traversal to perform weighted vector summation by attention or norm.

As for the traversal template, similarly to the discussion in Section~\ref{sec:graph_aware_loop}, we incorporate graph-semantic-aware loop transformation rules that allow \oursystem{} to leverage graph semantics to open up the trade-off between more data reuse opportunities and greater parallelism. As mentioned in Section~\ref{sec:two_templates}, initially, all statements are in the innermost loop in each instance derived from the traversal template. 
Loop hoisting is performed to enhance data reuse: The template features insertion points before and after the end of each loop level. For each statement, \oursystem{} finds the outermost level where it can be placed before applying the template. 
In addition, the template also provides a partial result aggregation method, which is applied during lowering by default, to reduce global memory traffic by accumulating results within a thread and within a warp before atomically adding them to the data in global memory.

\subsubsection{Operator Selection and Kernel Fusion}
Transformation and lowering passes at the inter-operator level need information about operator instances, specifically operator preference %
and the feasibility of kernel fusion.
Preference level is the mechanism \oursystem{} uses to select the operator instance when there are multiple candidates. For example, an operator instance derived from the GEMM template may %
 have an alternative derived from the traversal template but the alternative would lead to lower performance due to much lower data reuse. 
For good performance, operator instances derived from the GEMM template are assigned a higher preference level than those derived from the traversal template unless otherwise specified. Instances that fall back to PyTorch have the lowest preference level.

Operator instances also provide methods to determine the feasible operators to be fused within the IR. %
Operator instances derived from the GEMM template can be fused with the consumer if 1) the latter multiplies the row vectors in the GEMM output with scalars and 2) the two operators are in the same loop (nest). Operator instances derived from the traversal template can be fused with each other as long as they are in the same loop (nest).
If the inter-operator level pass finds that some temporary variables are created and merely used inside the fused operator, it passes that knowledge to the method so that the variable no longer needs to be created in the global memory.

\subsection{Backward Propagation}
\label{sec:bck_prop}

Similarly to PyTorch, \oursystem{} supports auto-differentiation by %
maintaining the backward propagation counterparts of the operators.
\oursystem{} first emits the backward propagation via inter-operator level IR, and removes unused gradients and their computation.
The lowering and code generation schemes are similar to those in forward propagation.
However, additional processing is needed because the PyTorch auto-differentiation requires the backward propagation methods to be paired with the forward propagation methods in the \texttt{autograd.Function} definitions. To achieve this, \oursystem{} bookkeeps the kernel calls in each forward propagation method. For each forward propagation method, \oursystem{} puts all the corresponding backward propagation kernel calls in the body of the backward propagation method.

\subsection{Code Generation}
\label{sec:code_gen}
The code generation procedure emits code based on the CUDA kernel specifications detailed in the form of intra-operator IR. Kernel code generation is fairly straightforward and is implemented using a template-based approach. 
\oursystem{} then emits the host functions that configure grids and blocks, gets raw pointers from the \texttt{libtorch} \texttt{at::Tensor} references, and launches the corresponding kernel. The host functions are exported via \texttt{pybind11} utilities. 

The \oursystem{} performs a pass that scans all the functions generated to collect a list of preprocessing required for the input dataset, involving transposition, converting COO to CSR, etc. The code generator then emits the preprocessing code.

\subsection{Applicability of the Optimizations to GNNs.}\label{sec:gnn_applicability}
Linear operator reordering and compact materialization are specific to RGNNs. Linear operator reordering is specific to RGNNs because RGNNs typically require linear projections from different semantic spaces, introduced by the heterogeneity of node types and edge types, to a common space before further operations. 
Compact materialization is specific to RGNNs because of the additional tensor dimension brought in by different node types and edge types. 

Some of the intra-operator IR optimizations could benefit ordinary GNNs, which can be treated as a special case of RGNNs whose relation type number is one. Intra-operator level IR allows specification of both data access schemes and schedules, thus allowing flexible code generation to accommodate different dense or sparse tensor layouts, a need that often arises from compact materialization. However, the ability to generate code for different data access schemes and schedules can be beneficial when compiling ordinary GNNs.

\section{Evaluation and Discussion}
\label{sec:eval}

We evaluate \oursystem{} with the following questions to answer.

\begin{enumerate}[Q1.]
\item How does the performance of \oursystem{} compare with state-of-the-art systems? How does \oursystem{} achieve it?
\item How much improvement do the two optimizations detailed in \cref{sec:materialization,sec:inter_op_opt}, compaction materialization and linear operator reordering, make? 
\item Any architectural insights for GPU for RGNNs?
\end{enumerate}

\begin{figure*}[!htbp]\captionsetup[subfigure]{font=small}
\centering
\subcaptionbox{Training time}
[.45\linewidth]{\includegraphics[scale=0.6]{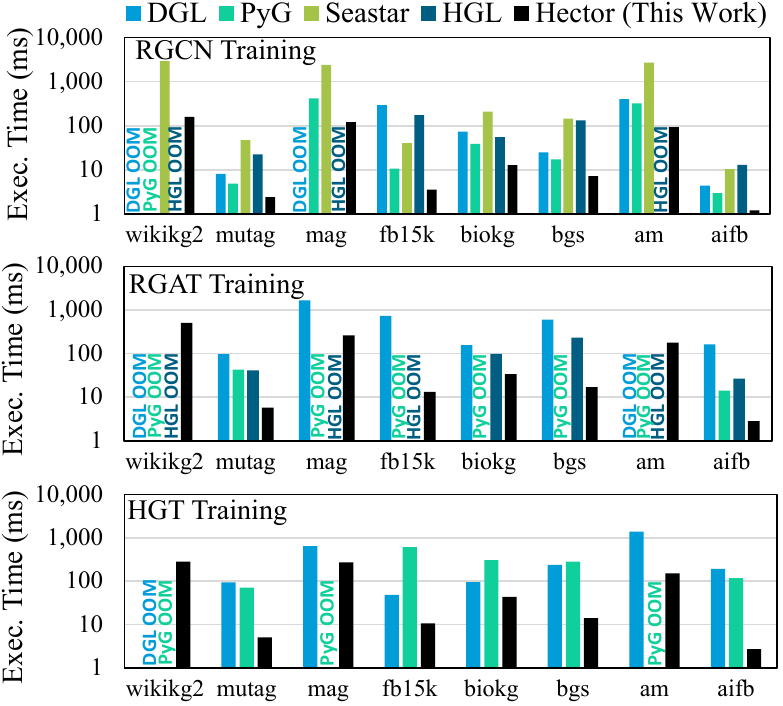}}
\subcaptionbox{Inference time}
[.45\linewidth]{\includegraphics[scale=0.6]{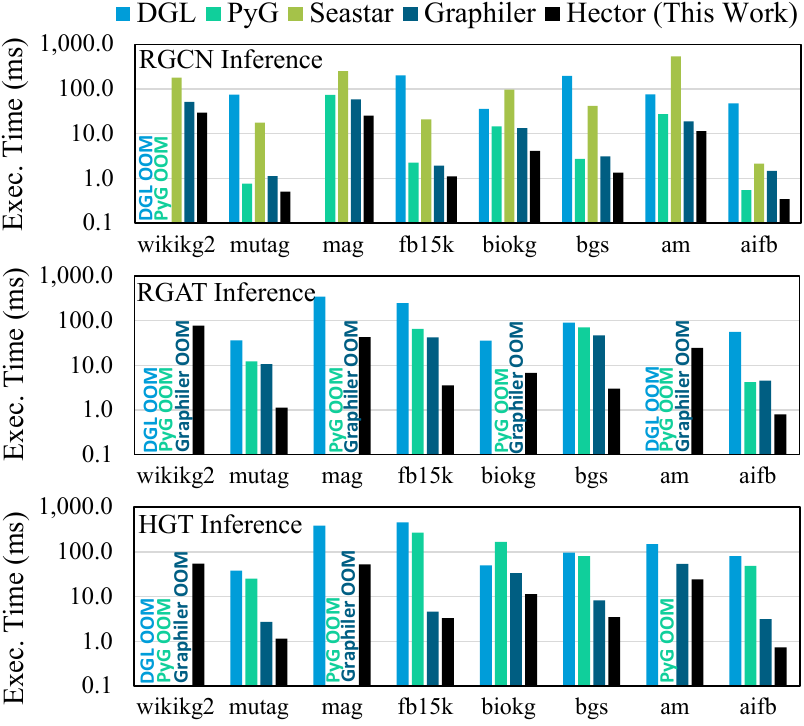}}
\caption{\label{fig:inference_results} Comparing the execution (Exec.) time of \oursystem{} best optimized code with previous work. Table~\ref{tab:datasets} shows the datasets used. }
\end{figure*}

\cref{sec:baseline_eval} answers Q1. \cref{sec:dse_eval} answers Q2, and further analyzes the performance implication of the two optimizations through a case study. \cref{sec:arch_analysis} addresses Q3.

\begin{table}[!htbp]
\centering
\caption{Heterogeneous graph datasets~\cite{aifb, mutag, bgs, am, huOpenGraphBenchmark2021, toutanovaObservedLatentFeatures2015} used in our evaluation. The numbers reflect the default preprocessing by the OGB and DGL packages, e.g., adding inverse edges.}\label{tab:datasets}
{\footnotesize
\begin{tabular}{lll|lll}
\toprule
\textbf{Name} & \multicolumn{1}{l}{\textbf{\begin{tabular}[c]{@{}l@{}}\#\ nodes\\ (\#\ types)\end{tabular}}} & \multicolumn{1}{l|}{\textbf{\begin{tabular}[c]{@{}l@{}}\#\ edges\\ (\#\ types)\end{tabular}}}  & \textbf{Name} & \multicolumn{1}{l}{\textbf{\begin{tabular}[c]{@{}l@{}}\#\ nodes\\ (\#\ types)\end{tabular}}} & \multicolumn{1}{l}{\textbf{\begin{tabular}[c]{@{}l@{}}\#\ edges\\ (\#\ types)\end{tabular}}}    \\ \midrule
aifb                           & 7.3K (7)    & 49K (104) & fb15k                           & 15K (1)  & 620K (474) \\ 
am                           & 1.9M (7) & 5.7M (108) & mag                           & 1.9M (4) & 21M (4) \\ 
bgs                              & 95K (27)   & 673K (122)&  mutag                          & 27K (5)   & 148K (50) \\     
biokg                           & 94K (5)   & 4.8M (51)&   wikikg2                     & 2.5M (1) & 16M (535) \\ \bottomrule 
\end{tabular}
}
\end{table}

\subsection{Methodology}
\label{sec:eval_methodology}

To assess performance, we measure the inference and training time of \oursystem{} and other systems on a single-GPU computer. Its hardware components include one Intel Core i9-9900K CPU, 128 GB dual-channel memory, and one Nvidia RTX 3090 GPU with 24GB memory. The operating system is Ubuntu 18.04.5, with kernel version 5.4.0-135. The CUDA and driver versions are 12.1 and 530.30.02, respectively. PyTorch and DGL versions are 2.0.1 and 1.1.1, respectively.

As shown in Table~\ref{tab:datasets}, we use public datasets from DGL~\cite{wang2019deep} and OGB~\cite{huOpenGraphBenchmark2021}.
We measure (1)~inference and (2)~training time on three RGNN models,  RGCN~\cite{rgcn}, RGAT~\cite{busbridge2019relational}, and HGT~\cite{hgt}, comparing with previous systems, involving DGL~\cite{wang2019deep}, PyG~\cite{fey2019fast}, Seastar~\cite{wuSeastarVertexcentricProgramming2021}, Graphiler~\cite{xieGraphilerCompilerGraph}, and HGL~\cite{guiHGLAcceleratingHeterogeneous}.
We ported Seastar artifacts to the same version of CUDA and Python packages as what \oursystem{} depends on because one of Seastar's dependencies, dgl 0.4, used an API deprecated since CUDA~11.

For RGCN, RGAT, and HGT, excluding comments, \oursystem{} took in 51 lines in total and produced more than 3K lines of CUDA kernel code, 5K lines of other C++ code to define host functions, and 2K lines of Python code to define subclasses of Pytorch \texttt{autograd.Function}. The implementation also involves 2K lines of Python code providing common utilities.

To best align with the hyper-parameters prior work used in its evaluation, we set the input and output feature dimensions as 64 and the number of heads as 1. 
We measure the inference and training time of the single layer used. 
In training, to obtain a loss, we compute the negative log-likelihood loss by comparing the output with a precomputed random label tensor. 
For each case, we run the full graph inference and training for at least 10 epochs and average the elapsed time.
To align with the existing system, nodes are presorted to enable segment MM for typed linear layers.

\subsection{Comparison with Prior Work}
\label{sec:baseline_eval}

For the performance of DGL and PyG, we measure all public implementations of these models from DGL, PyG, and Graphiler artifacts.
PyG provides two RGCN convolution layers: \texttt{RGCNConv} places nodes in segments of the same type but launches separate kernels for each of the node types, leading to device underutilization.
\texttt{FastRGCNConv} replicates weights and uses \texttt{bmm()}. It is consistently faster than the \texttt{RGCNConv} implementation.
Similarly, DGL's built-in segmentMM-based RGCN layer is faster than other DGL implementations.
For HGT, the DGL segmentMM-based \texttt{HGTConv} primitive generally has the best performance.
In the cases where some variants encounter OOM errors, we choose the best among those that run without issues.
Some cases are missing due to insufficient operator support, such as HGL on HGT and Graphiler on training. We do not measure HGL in inference because it is designed to optimize training.

Figure~\ref{fig:inference_results} shows that \oursystem{}'s best-optimized code consistently outperforms state-of-the-art systems. It achieves up to 9.9$\times$ speed-up in inference and up to 43.7$\times$ speed-up in training against the best of state-of-the-art systems. On geometric average, \oursystem{} gets 1.79$\times$, 8.56$\times$, 2.87$\times$ speed-up in inference via RGCN, RGAT, and HGT, respectively, and 2.59$\times$, 11.34$\times$, 8.02$\times$ speed-up in training RGCN, RGAT, and HGT, respectively. The performance advantage is larger in small graphs, demonstrating that \textbf{generating a single kernel that performs the computation across multiple edge types boosts the performance on small graphs %
compared to existing systems that run many small kernels}.

We see close performance achieved by Graphiler in RGCN and HGT inference. Graphiler leverages PyTorch utilities to produce TorchScript binaries before execution and utilizes edgewise parallelism for edgewise computation. Similarly to \texttt{RGCNConv}, it places node features into segments of the same type but runs separate kernels to perform a typed linear transformation. DGL and PyG under similar configurations achieve competitive performance. However, when it comes to RGAT, Graphiler suffers from performance degradation. Because Graphiler relies on pre-programmed fused kernels to deliver a significant portion of the performance boost~\cite{xieGraphilerCompilerGraph}, we postulate that the degradation is due to the non-exhaustiveness of these pre-programmed kernels~\cite{xieGraphilerRepositoryGithub2023}.
This reflects the drawbacks of compiler design without a code generation mechanism. By contrast, with two-level IR and a code generator, Hector achieves better performance, showing that \textbf{generating kernels with flexible access scheme that gather and scatter data on the fly eliminates redundant data movement and outperforms indexing/copying followed by hand-optimized GEMM and sparse kernels}. Besides, it is challenging to extend Graphiler's approach to training due to TorchScript's limited auto-differentiation support. For example, \texttt{dict} object creation is not supported, but it is a common way to express nodewise data and edgewise data. 

By comparing \oursystem{} with Seastar, which lowers all logic to sparse kernels, we realize that \textbf{sparse kernel code generation alone is not efficient in RGNNs: it is better to lower to GEMM kernels as much as possible}.

There are two reasons why \oursystem{} is more efficient in device memory usage. First, \oursystem{} only keeps a single copy of weights, as discussed in Section~\ref{sec:materialization}. Replicating weights also affects backward propagation because the gradient of each individual copy will be derived, occupying extra memory. Second, our compact materialization reduces memory and redundant computation, as explained in Section~\ref{sec:dse_eval}.

Notably, even without compact materialization or linear operator reordering, \oursystem{} still consistently outperforms existing systems, as Table~\ref{tab:base_vs_baseline} shows. In addition, the unoptimized \oursystem{} code triggers fewer OOMs than existing systems, with the only exception where the RGAT inference is run on mag and wikikg2. 
For comparison, we also show the statistics of the best optimized \oursystem{} code in Table~\ref{tab:base_vs_baseline}.

\begin{table}[!htbp]
\centering
\caption{Comparing to the best in state-of-the-art systems, speed-ups of \oursystem{} unoptimized (unopt.) code and that of \oursystem{} best optimized (b.\ opt.) code. Worst~(W), average~(M), and best~(B) cases. Numbers of OOMs \oursystem{} triggers~(\#E) are shown. \label{tab:base_vs_baseline}}
{\footnotesize
\begin{tabular}{lllllc|lllc}
 \toprule
&     & \multicolumn{4}{c}{\textbf{Training}}               & \multicolumn{4}{c}{\textbf{Inference}}              \\
&     & \multicolumn{1}{l}{\textbf{W}} & \textbf{M} & \textbf{B} & \multicolumn{1}{l}{\textbf{\#E}}  & \textbf{W} & \textbf{M} & \textbf{B} & \textbf{\#E}\\ \midrule
\multirow{3}{*}{\rotatebox[origin=c]{90}{\textbf{unopt.}}}& RGCN & 2.02 & 2.59 & 3.47  & \underline{0} & 1.51 & 1.79 & 2.19 & \underline{0} \\
& RGAT & 1.72 & 9.14 & 43.7 & \underline{2} & 1.41 & 5.02 & 9.89 & \underline{2} \\
& HGT  & 1.53 & 6.62 & 28.3 & \underline{0} & 1.20 & 1.90 & 4.31 & \underline{0} \\\hline
\multirow{3}{*}{\rotatebox[origin=c]{90}{\textbf{b.\ opt.}}}& RGCN & 2.02 & 2.76 & 3.48  & \underline{0} & 1.51 & 1.91 & 3.20 & \underline{0} \\
& RGAT & 4.61 & 11.3 & 55.4 & \underline{0} & 5.29 & 8.56 & 15.5 & \underline{0} \\
& HGT  & 2.17 & 8.02 & 43.1 & \underline{0} & 1.40 & 2.87 & 7.42 & \underline{0} \\
\bottomrule
\end{tabular}
}
\end{table}

\begin{figure}[!htbp]
\centering{\includegraphics[width=
\linewidth]{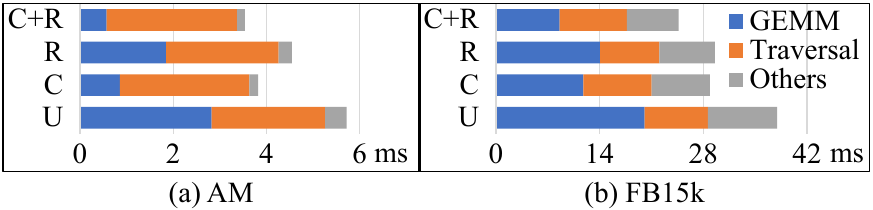}
}
\caption{\label{fig:perf_analysis} Breakdown of \oursystem{} RGAT inference on two datasets. Input and output dimensions are 64. Cases with compaction (C), linear operator reordering (R), and no optimization (U) are presented.}
\end{figure}

\begin{figure}[!htbp]
\centering{\includegraphics[width=\linewidth]{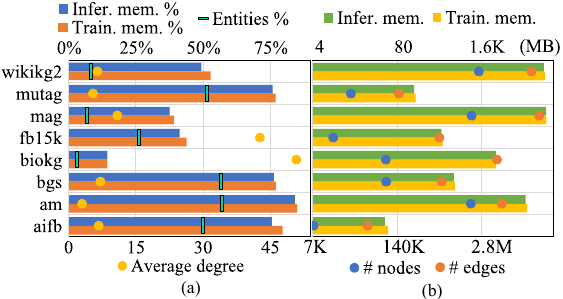}
}
\caption{\label{fig:dram_analysis} Memory usage when \oursystem{} runs training and inference on HGT. (b) shows the inference memory use (Infer.\ mem.) and training memory use (Train.\ mem.) of the unoptimized \oursystem{} code in MBs. (a) shows the portion of the memory use after applying compact materialization vs. the unoptimized \oursystem{} code. For comparison, the number of nodes (\#\ nodes), number of edges (\#\ edges), and average degree of datasets are shown as dot scatters. The entity compaction ratio of each dataset is also shown. Legend entries of each data series are placed next to the axis the series uses.}
\end{figure}

\begin{figure}[!htbp]
\centering{\includegraphics[width=\linewidth]{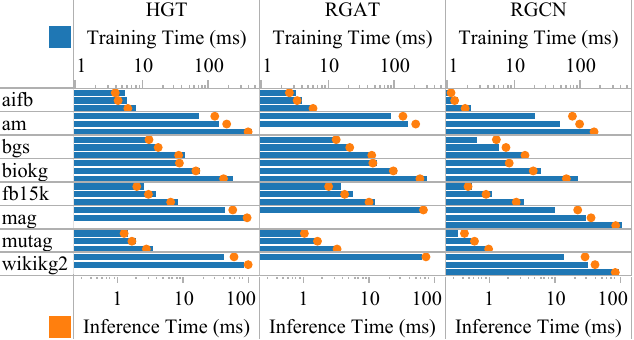}
}
\caption{\label{fig:perf_curve}\oursystem{} unoptimized performance. Each cell corresponds to one pair of dataset and model, where it is shown the time of (input dimension, output dimension) as (32, 32), (64, 64), and (128, 128) from the top to the bottom. Vacancy indicates OOM errors.}
\end{figure}

\subsection{Effects of Compact Materialization and Linear Operator Reordering}
\label{sec:dse_eval}
Now, we study the effects of compact materialization and linear operator reordering. They are detailed in \cref{sec:materialization,sec:inter_op_opt}. 
We investigate their effects on RGAT and HGT.

Table~\ref{tab:optimizations} shows the speed-up on top of \oursystem{} unoptimized code by these two optimizations. Due to compact materialization, when running RGAT on mag and wikikg2, \oursystem{} no longer triggers OOM errors. In addition, in some cases, the layout speeds up the execution as well due to the common subexpression elimination brought forth by the layout. Compact materialization is hardly possible without a code generation scheme or an IR design that decouples the model semantics, data layout, and operator-specific schedule.
Besides, \textbf{data layout choice, compact materialization in particular, allows further performance enhancement} while prior work usually focuses on the improvement of schedule given a specific sparse matrix format. This is shown by the great speedups in the ``C[ompact]'' columns in Table~\ref{tab:optimizations}.

To study how compact materialization reduces the memory footprint, We illustrate the \oursystem{} dram usage without compact materialization in Figure~\ref{fig:dram_analysis}(b) and the portion of dram usage with compact materialization in Figure~\ref{fig:dram_analysis}(a). For simplicity, we define the entity compaction ratio as the number of unique $(\text{source node}, \text{edge type})$ pairs divided by the number of edges. Figure~\ref{fig:dram_analysis}(b) shows that the memory use of inference and training is highly proportional to the number of edges of the datasets. Figure~\ref{fig:dram_analysis}(a) shows that compact materialization significantly reduces DRAM usage in all datasets. The memory footprint ratio of compact materialization compared with the memory footprint of the unoptimized code correlates with the entity compaction ratio. The memory footprint ratio is higher than the entity compaction ratio, as the memory footprint consists of edgewise data, nodewise data, and weights, whereas the compaction applies to edgewise data only. Besides, in case the average degrees are larger, the memory footprint ratio reduces more significantly, getting closer to the entity compaction ratio.

To better understand the performance benefits of optimizations, Figure~\ref{fig:perf_analysis} studies two cases. 
The entity compaction ratio of AM and FB15k are 57\% and 26\%, respectively. On AM, the time GEMM instances take is greatly reduced. By comparison, in FB15k, compaction brings less performance improvement due to the less significant GEMM reduction. 

In short, \textbf{due to the data-dependent nature of computation in RGNNs, there is no one-size-fits-all optimization strategy}. However, as shown in Table~\ref{tab:optimizations}, Enabling compaction and reordering obtains fairly good performance consistently and is the best fixed strategy on average in all four scenarios, i.e., $\{\text{RGAT}, \text{HGT}\}\times\{\text{training},\text{inference}\}$. If \oursystem{} presumably chooses the best configuration in every run, it could further get 1.06$\times$, 1.33$\times$, 1.02$\times$, and 1.08$\times$ speed-up in the four scenarios above, respectively. We leave autotuning to future work.

\begin{table}[htbp!]
\centering
\caption{Speed-up on top of \oursystem{} unoptimized code due to compaction~(C) and linear operator reordering~(R). Input and output dimensions are both 64. The highest speed-ups per task are in bold. }
\label{tab:optimizations}
{\footnotesize
\begin{tabular}{clccc|ccc} 
\toprule
\multicolumn{2}{l}{\multirow{2}{*}{\textbf{}}} & \multicolumn{3}{c}{\textbf{Training}}                                                                                          & \multicolumn{3}{c}{\textbf{Inference}}                                                                                          \\
\multicolumn{2}{l}{}                           & \multicolumn{1}{l}{\textbf{C}}           & \textbf{R}                               & \multicolumn{1}{c}{\textbf{C+R}}         & \textbf{C}                               & \textbf{R}                               & \textbf{C+R}                              \\ 
\midrule
\multirow{9}{*}{\rotatebox[origin=c]{90}{RGAT}}  & aifb    & \cellcolor[HTML]{D97460}0.80 & \cellcolor[HTML]{EEF4EC}\textbf{1.14}           & \cellcolor[HTML]{E08E7D}0.84          & \cellcolor[HTML]{FEFFFE}1.01          & \cellcolor[HTML]{E9F0E6}\textbf{1.19}           & \cellcolor[HTML]{F3F7F1}1.10          \\
 & am      & \cellcolor[HTML]{F2D1CB}0.94 & \cellcolor[HTML]{F2F6F0}\textbf{1.12}                    & \cellcolor[HTML]{FDF8F7}{0.99} & \cellcolor[HTML]{DAE6D5}1.31          & \cellcolor[HTML]{DEE8D9}1.28                    & \cellcolor[HTML]{C0D4B7}\textbf{1.54} \\
 & bgs     & \cellcolor[HTML]{F2D0C9}0.93 & \cellcolor[HTML]{EAF0E7}\textbf{1.18}           & \cellcolor[HTML]{FBFCFA}{1.04} & \cellcolor[HTML]{DDE8D9}1.29          & \cellcolor[HTML]{D7E4D2}{1.34}           & \cellcolor[HTML]{BCD1B3}\textbf{1.57} \\
 & biokg   & \cellcolor[HTML]{39771E}2.67 & \cellcolor[HTML]{E1EADD}1.26                    & \cellcolor[HTML]{38761D}\textbf{2.68} & \cellcolor[HTML]{38761D}\textbf{3.76} & \cellcolor[HTML]{D0DFC9}1.40                    & \cellcolor[HTML]{38761D}{3.74} \\
& fb15k & \cellcolor[HTML]{E8EFE5}1.20 & \cellcolor[HTML]{E7EFE4}1.20 & \cellcolor[HTML]{E0EADB}\textbf{1.27} & \cellcolor[HTML]{C5D7BD}1.50 & \cellcolor[HTML]{E1EADD}1.26 & \cellcolor[HTML]{B6CDAC}\textbf{1.62} \\
 & mag     & \cellcolor[HTML]{C3D6BB}1.51 & \multicolumn{1}{l}{\cellcolor[HTML]{F9FBF8}OOM} & \cellcolor[HTML]{BCD1B3}\textbf{1.57} & \cellcolor[HTML]{FFFFFF}1.00*          & \multicolumn{1}{l}{\cellcolor[HTML]{F3F7F2}OOM} & \cellcolor[HTML]{F8FAF7}\textbf{1.07} \\
 & mutag   & \cellcolor[HTML]{CC4125}0.70 & \cellcolor[HTML]{EFF4ED}\textbf{1.14}           & \cellcolor[HTML]{CC4125}0.73          & \cellcolor[HTML]{E4EDE0}1.23          & \cellcolor[HTML]{E3ECE0}{1.24}           & \cellcolor[HTML]{D5E2CF}\textbf{1.36} \\
 & wikikg2 & \cellcolor[HTML]{F4F8F3}1.09 & \multicolumn{1}{l}{\cellcolor[HTML]{F7FAF6}OOM} & \cellcolor[HTML]{F1F6EF}\textbf{1.12} & \cellcolor[HTML]{FFFFFF}1.00*          & \multicolumn{1}{l}{\cellcolor[HTML]{EDF3EB}OOM} & \cellcolor[HTML]{FEFEFD}\textbf{1.02} \\
 & AVERAGE & \cellcolor[HTML]{F0F5EE}1.13 & \cellcolor[HTML]{EBF1E8}1.17                    & \cellcolor[HTML]{EAF1E7}\textbf{1.18} & \cellcolor[HTML]{D5E2CF}1.36          & \cellcolor[HTML]{DEE8D9}1.28                    & \cellcolor[HTML]{C6D8BE}\textbf{1.49}
                  \\ 
\hline
\multirow{9}{*}{\rotatebox[origin=c]{90}{HGT}} & aifb    & \cellcolor[HTML]{F9E8E5}0.97 & \cellcolor[HTML]{C2D5B9}\textbf{1.52} & \cellcolor[HTML]{D1DFCA}1.40          & \cellcolor[HTML]{F0C9C1}0.92 & \cellcolor[HTML]{90B381}\textbf{1.94} & \cellcolor[HTML]{BBD0B1}1.58          \\
& am      & \cellcolor[HTML]{FAFCF9}1.05 & \cellcolor[HTML]{F1F6EF}1.12          & \cellcolor[HTML]{E9F0E6}\textbf{1.19} & \cellcolor[HTML]{F9FBF8}1.06 & \cellcolor[HTML]{DAE6D5}1.32          & \cellcolor[HTML]{CEDDC7}\textbf{1.42} \\
& bgs     & \cellcolor[HTML]{FEFCFB}1.00* & \cellcolor[HTML]{F2F6F0}1.11          & \cellcolor[HTML]{EAF1E8}\textbf{1.18} & \cellcolor[HTML]{F3D5CF}0.94 & \cellcolor[HTML]{E1EBDD}\textbf{1.25} & \cellcolor[HTML]{E3ECDF}1.24          \\
& biokg   & \cellcolor[HTML]{D6E3D1}1.35 & \cellcolor[HTML]{FCFDFB}1.03          & \cellcolor[HTML]{CFDEC8}\textbf{1.41} & \cellcolor[HTML]{CBDBC4}1.45 & \cellcolor[HTML]{F8FAF6}1.07          & \cellcolor[HTML]{BBD0B2}\textbf{1.58} \\
& fb15k   & \cellcolor[HTML]{E7A79A}0.88 & \cellcolor[HTML]{F2F6F0}\textbf{1.11} & \cellcolor[HTML]{F8E5E2}0.96          & \cellcolor[HTML]{D35E46}0.77 & \cellcolor[HTML]{ECF2EA}\textbf{1.16} & \cellcolor[HTML]{E59F91}0.86          \\
& mag     & \cellcolor[HTML]{E3ECDF}1.24 & \cellcolor[HTML]{F9FBF8}1.06          & \cellcolor[HTML]{D7E4D2}\textbf{1.34} & \cellcolor[HTML]{C9DAC1}1.46 & \cellcolor[HTML]{F3F7F2}1.10          & \cellcolor[HTML]{AAC59F}\textbf{1.72} \\
& mutag   & \cellcolor[HTML]{FEFDFD}1.00 & \cellcolor[HTML]{D9E5D4}\textbf{1.32} & \cellcolor[HTML]{DAE6D5}1.32          & \cellcolor[HTML]{F4D7D1}0.94 & \cellcolor[HTML]{AFC8A4}\textbf{1.68} & \cellcolor[HTML]{C4D6BC}1.50          \\
& wikikg2 & \cellcolor[HTML]{E6EEE2}1.22 & \cellcolor[HTML]{F7FAF6}1.07          & \cellcolor[HTML]{D8E5D3}\textbf{1.33} & \cellcolor[HTML]{E1EBDD}1.26 & \cellcolor[HTML]{EDF3EB}1.15          & \cellcolor[HTML]{C3D6BB}\textbf{1.51} \\
& AVERAGE & \cellcolor[HTML]{F6F9F5}1.08 & \cellcolor[HTML]{ECF2EA}1.16          & \cellcolor[HTML]{E1EADD}\textbf{1.26} & \cellcolor[HTML]{F7F9F5}1.07 & \cellcolor[HTML]{DBE6D6}1.31          & \cellcolor[HTML]{D0DFCA}\textbf{1.40}
        \\
\bottomrule
\end{tabular}
}
\begin{flushleft} \footnotesize *Normalized by the performance with compact materialization (C) because the unoptimized version triggers OOM errors. 
\end{flushleft}
\end{table}

\subsection{Analyzing the Architectural Characteristics}
\label{sec:arch_analysis}

\begin{figure}[!htbp]
\centering{\includegraphics[width=\linewidth]{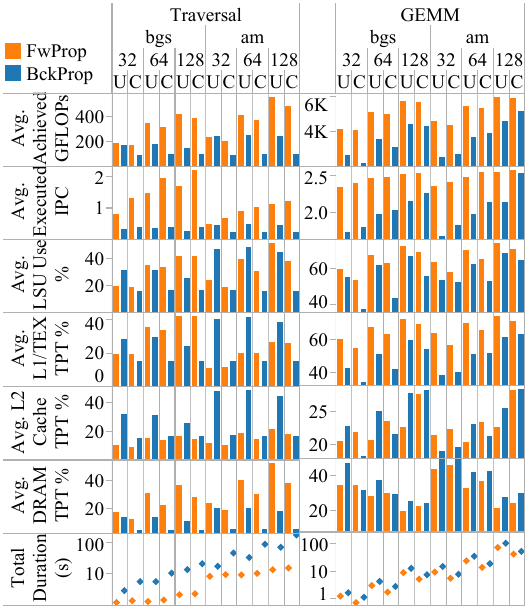}}
\caption{\label{fig:arch_number}Architectural metrics of \oursystem{} kernels in the forward~(Fw) and backward~(Bck) propagation when running \oursystem{} on RGAT with compaction (C) and without (U). For each kernel category, aggregated duration and average (Avg.) metrics, e.g., instructions per cycle (IPC) and various throughputs (TPT), are reported.}
\end{figure}

We show the average time of unoptimized \oursystem{} in Figure~\ref{fig:perf_curve}. We also further profile generated kernels when running \oursystem{} on RGAT on bgs and am, as shown in Figure~\ref{fig:arch_number}.

One thing to note is the sublinear time increase in Figure~\ref{fig:perf_curve}: when the input and output dimension doubles, the amount of computation and memory accesses becomes close to 4$\times$ those of the original, but the time increase is typically lower than 2$\times$ of the original. The reason is increased computation throughput when the size increases, as corroborated by Figure~\ref{fig:arch_number}. Moreover, we observed higher throughput when the graph scale increases, e.g., from bgs to am in Figure~\ref{fig:arch_number}. Similarly, we witnessed the cuBLAS throughput increases steadily when we keep the right matrix size as (64, 64) and increase the number of rows of the left matrix from 1M (2\textsuperscript{17}) to 8M (2\textsuperscript{20}). These suggest that \textbf{an RGNN system should be memory-efficient in order to accommodate larger models and datasets to fully utilize the massive resources on GPUs}. By eliminating unnecessary data copies, \oursystem{} achieves better memory efficiency than state-of-the-art systems.

The instruction per cycle (IPC) charts in Figure~\ref{fig:arch_number} indicate the traversal kernels are generally latency-bound: on RTX 3090, IPC is ideally 4 as each streaming multiprocessor (SM) has four schedulers. Backward propagation kernels have lower throughput due to worsened latency and increased memory bandwidth consumption by doubled memory accesses compared to forward propagation. In backward propagation, backward traversal kernels compute gradients using atomic updates, therefore hindering the throughput; GEMM kernels also on average have lower performance due to outer products that compute the delta of weights.

\section{Related Work}
\label{sec:related_work}

\noindent\textbf{General GPU-accelerated GNN libraries.} DGL~\cite{wang2019deep} and PyG~\cite{fey2019fast} are among the most popular GNN Python packages that enable easy development and evaluation of GNN models.  DGL~\cite{wang2019deep} proposes to implement GNN as SpMM/SDDMM operations. PyG's key scheme is scatter and gather operations that switch between edge-parallel regions and node-parallel regions. \oursystem{} instead built upon GEMM and traversal templates. By lowering the operators to GEMM as much as possible, \oursystem{} obtains better RGNN performance.
Besides,  DGL, PyG and work based on them do not currently provide inter-operator level IR. Our work shows the benefit of capturing
inter-operator and inter-relation opportunities, e.g., linear-operator reordering, by operator rewrite at the inter-operator level IR. Systems without IR at this level eagerly execute operators without support for such optimizations.

\noindent
\textbf{GNN end-to-end compilers.} Seastar~\cite{wuSeastarVertexcentricProgramming2021} proposes a vertex-centric compiler stack to generate performant kernels throughout the training and/or inference of the model. Graphiler~\cite{xieGraphilerCompilerGraph} proposes to program the message passing data flow graph and devises several TorchScript transforms to emit highly optimized inference code. Similarly, HGL~\cite{guiHGLAcceleratingHeterogeneous} is an RGNN compiler. These prior arts 1) expose PyTorch tensors as operands of all operations to users and 2) replicate weight to unleash parallelism due to a lack of support for flexible data access schemes and/or code generation. Thus, they suffer more or less from memory inefficiency and performance degradation.
Although the general concept of multi-level IR is not new, this work proposes new optimizations appropriate for each level and effective in reducing data movement and code bloat in the current state of practice:
Linear operator reordering and compact materialization are two key and novel features to capture and eliminate repetitive computation across edge types. Section~\ref{sec:gnn_applicability} discussed the generalizability of this work.

\noindent
\textbf{Kernel code optimization.} FeatGraph~\cite{huFeatGraphFlexibleEfficient2020a} proposes code optimization framework on top of TVM~\cite{chenTVMAutomatedEndtoEnd2018} for user-defined-function-enabled SpMM and SDDMM. Some work proposed optimization  for specific GNNs kernels. GE-SpMM~\cite{huangEfficientSparseMatrix2021, huangGESpMMGeneralPurposeSparse2020}, and work~\cite{hidayetogluAtScaleSparseDeep2020} propose optimized schedules for SpMM. Others involve Seastar~\cite{wuSeastarVertexcentricProgramming2021}, PyTorch-Direct~\cite{minLargeGraphConvolutional2021}, and TLPGNN~\cite{fuTLPGNNLightweightTwoLevel2022}. As our work shows, SpMM/SDDMM is not the only essential kernel in end-to-end RGNN execution. And our work is orthogonal to these prior arts as they can be incorporated into \oursystem{} as operator-specific schedules or new templates.

\noindent
\textbf{Code generation.} 
SparseTIR~\cite{yeSparseTIRComposableAbstractions2022} and TACO~\cite{kjolstadTensorAlgebraCompiler2017} propose IR and code generator for sparse tensor operations.
MLIR~\cite{lattnerMLIRScalingCompiler2021} proposes multi-level IR design for deep learning.
Aligned with this direction, FusedMM~\cite{rahmanFusedMMUnifiedSDDMMSpMM2021} unifies the SpMM and SDDMM CPU kernel. 
\oursystem{} is different as a higher-level compiler that optimizes the type of operators and generates efficient kernels to handle multiple edge/node types in the RGNN execution. SparseTIR and TACO are tensor-level compilers for sparse operators that may or may not specialize in deep learning.
While we do not intend to reinvent the general-purpose sparse tensor code generator for completeness or performance, some of these works inspire us and may be incorporated to enhance the \oursystem{} code generator.

\section{Discussion and Future Work}\label{sec:future_work}

\noindent
\textbf{Extending the work to support for new optimizations.} \oursystem{} is designed as an extensible framework to prototype and evaluate new techniques. First, inter-operator optimizations can be prototyped as inter-operator level passes. Second, data layout optimizations can be supported by adding the corresponding intermediate data and adjacency access schemes discussed in Section~\ref{sec:inter_op_ir}.
We plan to explore 1) if different sparse formats have any impact on the sparse kernel performance and 2) if there is any optimization opportunity in changing the intermediate data layout. E.g., storing edge attention and edge message together in one tensor where the innermost dimension becomes (size of hidden dimension + 1) would further reduce the number of kernels launched.
Third, kernel optimizations can be prototyped as a kernel template and operator instances based on it. Alternatively, they can be implemented as operator-specific schedules.

\noindent
\textbf{Extending the work to use it in Distributed Systems.}
Due to the page limit, we focused this work on single-GPU performance. The kernels \oursystem{} generated could serve distributed systems, e.g., DistDGL~\cite{zhengDistDGLDistributedGraph2020}. Since performance improvement results from the reduction of data movements and memory footprints, it also applies to distributed systems.

\noindent
\textbf{Devise algorithms to select the layouts, optimizations, and schedules according to model, input graph, and GPU architecture.} One of the most important compiler research problems is the algorithm that makes choices among the candidates in the design space. Apart from the input graph, the specific microarchitecture of each GPU model also makes a difference due to the architecture-specific features available, e.g., asynchronous loading to shared memory since Ampere~\cite{nvidiaControllingDataMovement2020}, and different microarchitecture characteristics in each model. Therefore, it is meaningful to investigate their impact and incorporate them into decision making.

\noindent
\textbf{Optimize data movement in minibatch training.} Graphs not fiting into GPU memory must stay in host memory or even storage during RGNN execution. In each step, the subgraphs are sampled and transferred to the GPU. With knowledge of graph semantics, data layout, and operator-specific schedules, \oursystem{} can help improve the scheduling of sampling and data transfer and generate CUDA kernels that gather data from the host memory on the fly~\cite{minLargeGraphConvolutional2021}.

\noindent
\textbf{Incorporate TACO to enhance the traversal code generation.} In this work, we craft the code generators on our own for quick prototyping and focus on high-level optimizations. As this work establishes our understanding of what constructs are needed in the code generators for the traversal kernels, we think it viable to incorporate TACO for the code generation in the future because TACO provides a mature compiler infrastructure that enables the expression and application of optimizations~\cite{kjolstadTensorAlgebraCompilation2019} for sparse tensor operations in a principled manner, e.g., loop transformations. However, RGNN scenarios still pose several open challenges to TACO, especially in edge-centric operations. Take the edge-wise dot product when computing $att_{HGT}$ in Figure~\ref{fig:rgat_layer} as an example. First, to balance the workload, we evenly split the edgewise loop and assign them to threading blocks. If we specify the source-node-wise loop and destination-node-wise loop as two dimensions in the TACO iteration space, we need to fuse these two loop levels to form the edgewise loop to split, but such loop fusion between two loop levels of iteration variables is not supported by TACO yet. Alternatively, we can specify the edgewise loop index as one dimension in the iteration space. In this case, we need indirect addressing to retrieve node data: We need to retrieve \textcircled{1} the source/destination node index by edgewise loop index and then \textcircled{2} the node data. However, indirect addressing is not natively supported in TACO and thus poses the second challenge.

\section{Conclusion}\label{sec:conclusion}
RGNN execution has performance challenges due to the inherent computation pattern, the gap between the programming interface and kernel APIs, and the high kernel optimization cost due to the kernels' coupling with layout and heterogeneity.
To systematically address these challenges, we presented the \oursystem{} IR and code generator for end-to-end RGNN training and inference. The IR design \textit{decouples} the model semantics, data layout, and operator-specific schedule, and \textit{expresses} these opportunities to allow them to be integrated into the design space as integral elements. 
Based on a GEMM template and a traversal template, \oursystem{} already achieves up to 43.7$\times$ speed-up in training and inference compared to state-of-the-art systems. Linear operator reordering and compact tensor materialization obtain up to 3.8$\times$ speed-up compared to the \oursystem{} unoptimized code.

\section*{Acknowledgments}
This work was partially supported by a grant from HPE, a gift from AMD/Xilinx, and equipment gifts from AMD/Xilinx and NVIDIA.
The authors thank anonymous reviewers for their constructive comments.
Kun owes thanks to Dr. Seung Won Min and Prof. Guohao Dai for initiating this collaboration and precious advice on this project.
The authors thank Prof. Deming Chen, Dr. Penporn Koanantakool, Dr. Dejan Milojicic, and Prof. Vikram Adve for hosting meetings and seminars where many constructive feedbacks were given.

\bibliographystyle{plain}
\bibliography{sections/ref,sections/final_kuns_manually_revised_based_on_zotero_generated}

\end{document}